\makeatletter
\@ifundefined{@parse@version@dash}{%
\def\@parse@version#1{\@parse@version@0#1}
\def\@parse@version@#1/#2/#3#4#5\@nil{%
\@parse@version@dash#1-#2-#3#4\@nil}
\def\@parse@version@dash#1-#2-#3#4#5\@nil{%
  \if\relax#2\relax\else#1\fi#2#3#4 }
}{}
\makeatother  

\documentclass[prl,twocolumn,superscriptaddress,showpacs,floatfix,longbibliography]{revtex4-2}
\usepackage{mathrsfs,braket}
\usepackage{amssymb, amsbsy, amsmath, latexsym, dsfont, array, layout, graphicx,mathrsfs,color,ulem,bm}
\usepackage[colorlinks=true,citecolor=blue,urlcolor=blue,linkcolor=blue]{hyperref}

\begin{document}

\title{Dynamic Signatures of Non-Hermitian Skin Effect and Topology in Ultracold Atoms}

\author{Qian Liang}
\thanks{These authors contributed equally to this work}
\author{Dizhou Xie}
\thanks{These authors contributed equally to this work}
\author{Zhaoli Dong}
\thanks{These authors contributed equally to this work}
\affiliation{%
Interdisciplinary Center of Quantum Information, State Key Laboratory of Modern Optical Instrumentation, Zhejiang Province Key Laboratory of Quantum Technology and Device, Department of Physics, Zhejiang University, Hangzhou 310027, China
}%
\author{Haowei Li}
\affiliation{CAS Key Laboratory of Quantum Information, University of Science and Technology of China, Hefei 230026, China}
\affiliation{CAS Center For Excellence in Quantum Information and Quantum Physics, Hefei 230026, China}
\author{Hang Li}
\affiliation{%
Interdisciplinary Center of Quantum Information, State Key Laboratory of Modern Optical Instrumentation, Zhejiang Province Key Laboratory of Quantum Technology and Device, Department of Physics, Zhejiang University, Hangzhou 310027, China
}%
\author{Bryce Gadway}
\affiliation{Department of Physics, University of Illinois at Urbana-Champaign, Urbana, IL 61801-3080, USA}
\author{Wei Yi}
\email{wyiz@ustc.edu.cn}
\affiliation{CAS Key Laboratory of Quantum Information, University of Science and Technology of China, Hefei 230026, China}
\affiliation{CAS Center For Excellence in Quantum Information and Quantum Physics, Hefei 230026, China}
\author{Bo Yan}
\email{yanbohang@zju.edu.cn}
\affiliation{%
Interdisciplinary Center of Quantum Information, State Key Laboratory of Modern Optical Instrumentation, Zhejiang Province Key Laboratory of Quantum Technology and Device, Department of Physics, Zhejiang University, Hangzhou 310027, China
}%

\begin{abstract}
The non-Hermitian skin effect (NHSE),  the accumulation of eigen wavefunctions at boundaries of open systems, underlies a variety of exotic properties that defy conventional wisdom.
While NHSE and its intriguing impact on  band topology and dynamics have been observed in classical or photonic systems, their demonstration in a quantum gas system remains elusive. Here we report the
experimental realization of a dissipative Aharonov-Bohm chain---a non-Hermitian topological model with NHSE---in the momentum space of a two-component Bose-Einstein condensate. We identify signatures of
NHSE in the condensate dynamics, and perform Bragg spectroscopy to resolve topological edge states against a background of localized bulk states. Our work sets the stage for further investigation
on the interplay of many-body statistics and interactions with NHSE, and is a significant step forward in the quantum control and simulation of non-Hermitian physics.
\end{abstract}

\maketitle

A quantum system coupled to its environment generally suffers from particle loss and decoherence, thus exhibiting qualitatively different phenomena from an isolated conservative system. Much insight can be
obtained of these open quantum systems by resorting to a non-Hermitian description~\cite{QJ,uedareview}, under which a wealth of intriguing features naturally emerge, including parity-time symmetry and spectral
singularity~\cite{ptbender,ptreview1,ptreview2,ptreview3}, non-reciprocal and chiral transport~\cite{nonrecNature}, as well as non-Hermitian topology and unconventional band theory~\cite{nonHtopo1,nonHtopo2,WZ1,murakami}.
Within this context, a particularly fascinating phenomenon is the recently discovered non-Hermitian skin effect (NHSE)~\cite{WZ1,murakami,WZ2,ThomalePRB,Budich,mcdonald,alvarez,fangchenskin,kawabataskin,stefano,yzsgbz,tianshu,lli,Zhou2021,Guo2021}. Through what transpires as the nominal bulk eigenstates undergo exponential localization at the open boundaries, the NHSE fundamentally reshapes spectral, band, and dynamic properties of an open system, necessitating a
non-Bloch band theory to account for the non-Hermitian topology~\cite{WZ1,murakami,WZ2}, and leaving signatures in the dynamics either driven by a non-Hermitian effective Hamiltonian~\cite{stefano,tianyuquench}
or under the master equation~\cite{wzopen,stefanoopen}. While the recent observation of NHSE and its rich consequences have stimulated intense interest~\cite{teskin,photonskin,XDW+21,metaskin,teskin2d,scienceskin},
its experimental implementation in quantum gases remains unexplored.

In this work, we experimentally realize a non-Hermitian topological Hamiltonian with NHSE in a momentum lattice of ultracold atoms~\cite{Meier2016, Meier2018,Lapp2019,Xie2019}. Specifically, we engineer a chain of coupled Aharonov-Bohm (AB) rings along a synthetic momentum lattice with a Bose-Einstein condensate of $^{87}$Rb atoms~\cite{Xie2018,yanring,Chen2021}. We experimentally implement a chain of five unit cells with open boundary condition (OBC), and identify signatures of the NHSE through the condensate dynamics. {In particular, our measurements confirm the presence of a directional bulk flow, which drives eigenstates toward boundaries under the OBC and has been identified as the origin of the NHSE~\cite{stefano,fangchenskin,kawabataskin}.
For our finite-size system, the bulk dynamics is realized by initializing the condensate in a central site and probing the dynamics before atoms propagate to the boundary.}
In contrast, to resolve topological edge states amongst the similarly
localized bulk modes, we perform the Bragg spectroscopy at the edge of the AB chain to probe its eigenspectrum.
{The spectroscopy reveals the emergence of in-gap edge states as the parameters are tuned, consistent with the theoretically predicted non-Bloch topological phase transition.}
Given their flexible tunability and many-body nature, cold atomic gases offer an enticing platform wherein the impact of NHSE on a wide variety of physical phenomena can be systematically explored in the future.

\begin{figure}[tbp]
\centering
\includegraphics[width= 0.48\textwidth]{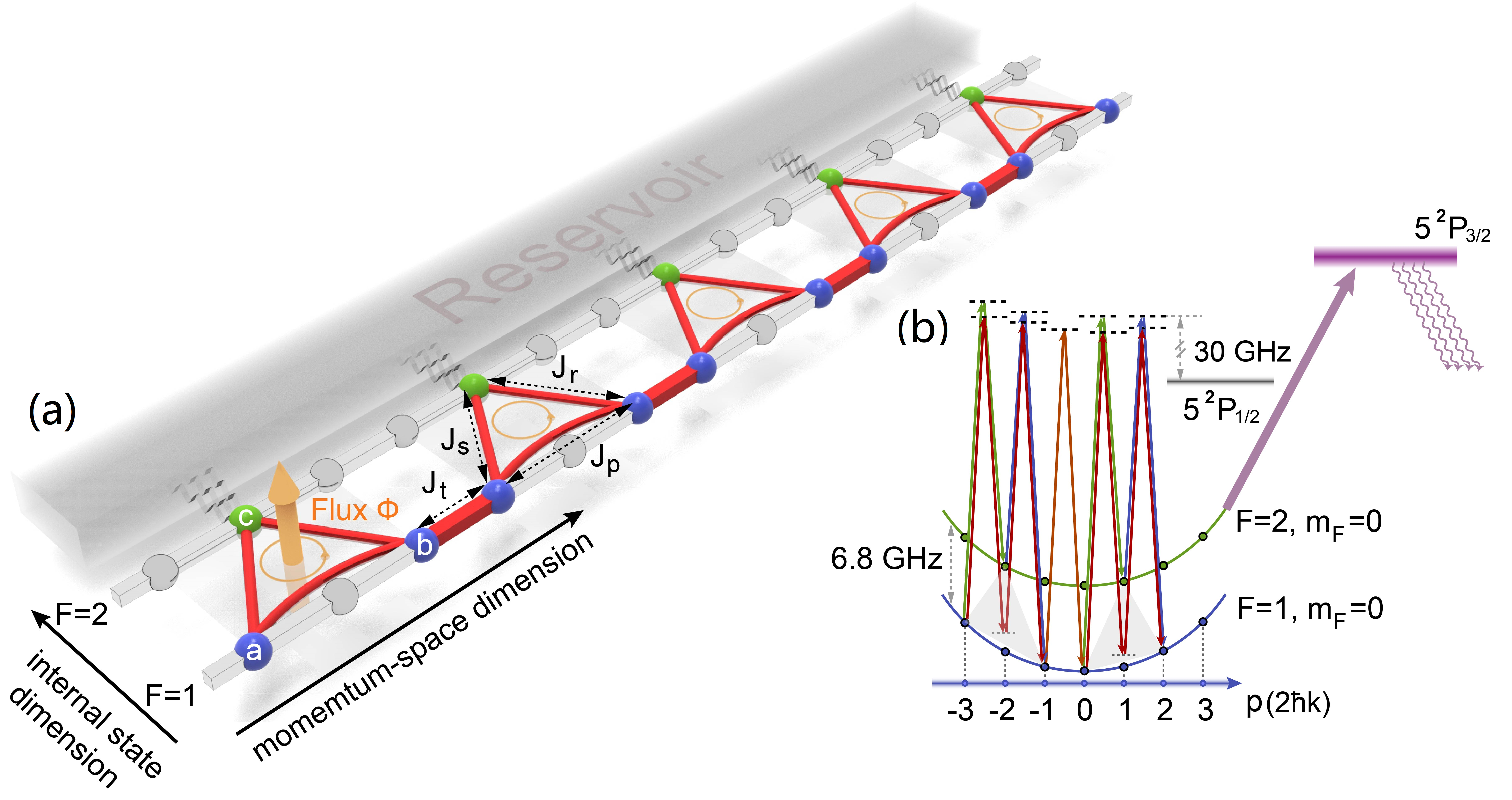}
\caption{\label{fig:model}
Topological Aharonov-Bohm chain with NHSE. (a) Schematic illustration of the  Aharonov-Bohm chain. Each unit cell includes three sites ($a,~b,~c$), forming a closed triangular loop. Site $c$ is further coupled to a reservoir that introduces dissipation to the system. While hopping rates $J_{p,t,r,s}$ are induced by Raman or Bragg processes, their phases contribute to a synthetic magnetic flux $\phi$ through each ring.
(b) The momentum lattice in (a) is generated by multi-frequency Raman and Bragg processes that couple discrete momentum and hyperfine states (both with $m_F=0$) in the $F=1$ and $F=2$ manifolds of $^{87}$Rb atoms. Within each unit cell, sites $(a,~c)$ and $(b,~c)$ are coupled through resonant Raman processes, with a momentum difference $2\hbar k$; while sites $(a,~b)$ are coupled through a four-photon resonant Bragg process, with a momentum difference of $4\hbar k$. Here $k$ is the wave vector of a $795$nm laser.
Subsequently, in the $n$th unit cell ($n\in\mathbb{Z}$), sites $a$ and $b$ represent the momentum states $|p=3n\rangle$ and $|p=3n+2\rangle$ (in units of $2\hbar k$), respectively, of the hyperfine ground state $|F=1,m_F=0\rangle$ in the $^5S_{1/2}$ manifold. Site $c$ is encoded into the momentum state { $|p=3n+1\rangle$} of the hyperfine ground state $|F=2,m_F=0\rangle$.
Dissipation is induced by a near-resonant coupling of atoms on site $c$ to the $^5P_{{3}/{2}}$ manifold.
}
\end{figure}

{\it Experimental implementation.}
Our experimental setup is illustrated in Fig.~\ref{fig:model}. Each unit cell of the AB chain consists of a triangular loop of three sublattice sites $(a,~b,~c)$, which are encoded in the combined synthetic dimensions of the ground-state hyperfine states
($|F=1,m_F=0\rangle$ for sites $a,~b$, and $|F=2,m_F=0\rangle$ for site $c$) and atomic momentum. While adjacent sites are coupled via Raman or Bragg transitions [see Fig.~\ref{fig:model}(b)], a synthetic magnetic
flux (denoted as $\phi$) through the rings can be generated and controlled through the phases of the coupling lasers. Dissipation is introduced at the vertex of the unit cells, as atoms on site $c$ are state-selectively coupled to electronically excited states in the $5{}^2P_{3/2}$ manifold, and subsequently lost from the system through spontaneous emission.

For atoms that remain in the dissipative AB chain, their dynamics is effectively driven by a non-Hermitian Hamiltonian~\cite{supp}
\begin{align}\label{Heff2}
	H &= \sum_n(\Delta-i\gamma)c^{\dagger}_n c_n +\sum_n[(J_p b^\dagger_n a_n + J_t a^\dagger_{n+1}b_n \nonumber \\
	&+ J_se^{i\phi}c^\dagger_n a_n +J_rc^\dagger_n b_n) + \mathrm{H.c.}],
\end{align}
where $a^{\dagger}_n (a_n)$, $b_n^{\dagger} (b_n)$ and $c_n^{\dagger} (c_n)$ are respectively the creation (annihilation) operators for sublattice sites $a$,$b$ and $c$ of the $n$th unit cell.
In the following, we denote the corresponding states as $|n,a\rangle$, $|n,b\rangle$, and $|n,c\rangle$, respectively. The laser-induced hopping rates $J_{p,t,r,s}$ between adjacent sites are illustrated in Fig.~\ref{fig:model}(a),
where the phase $\phi$ corresponds to the synthetic magnetic flux through the ring. The laser-induced dissipation on site $c$ is characterized by the effective {loss term} $\gamma$, and an on-site energy
offset $\Delta$ is present if the Raman-Bragg coupling to site $c$ is detuned. All these parameters are easily tunable, thanks to the flexible control afforded by the momentum-lattice engineering~\cite{supp}.

In each experiment, we start with a  $^{87}$Rb Bose-Einstein condensate in an optical dipole trap with a typical atom number of  $1\times 10^5$. The dipole trap frequencies are $\sim 2\pi\times(40, 100,115)$\ Hz. The Raman-Bragg lasers are imposed along the weak-trapping direction to construct the momentum lattice~\cite{supp}. For detection, we turn off the optical dipole trap and all the Raman-Bragg lasers, and take an absorption image after a $20$\ ms time-of-flight. In this way, atoms in different momentum states are separated and we can extract the atom population in different momentum states.

\begin{figure}[tbp]
\centering
\includegraphics[width= 0.48\textwidth]{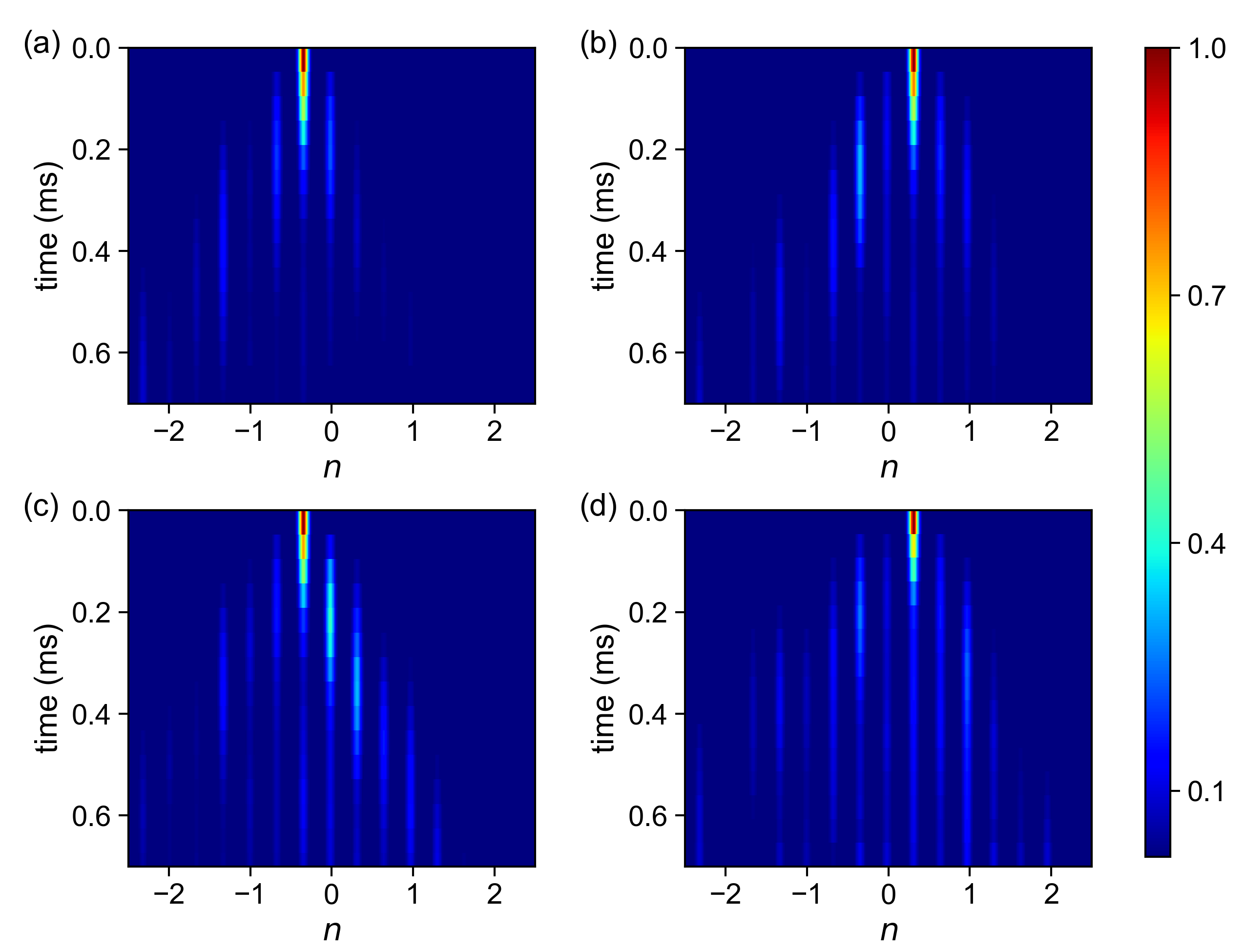}
\caption{\label{fig:typical}
Bulk dynamics along the momentum lattice, where we show the normalized atom population for each sublattice site.
(a)(b) are the non-Hermitian cases with $\gamma =h\times 1.3(2)$\ kHz; (c)(d) are the Hermitian cases with $\gamma =0$.
Atoms are initially prepared in the state $|n=0,a\rangle$ in (a)(c), and $|n=0,b\rangle$ in (b)(d).
The color bar indicates the atom population in each sublattice site, normalized by the initial atom number and shown in the order
$|n,a\rangle,|n,c\rangle, |n,b\rangle$ (from left to right) for a given unit cell $n$.
The experimental parameters are $\{J_s,~J_r,~J_p,~J_t\} = h \times \{1.07(2),~1.07(1),~0.65(3),~0.81(1)\}$ kHz, $\phi = -\pi/2$, and $\Delta = 0$.
}
\end{figure}

\begin{figure*}[tbp]
\centering
\includegraphics[width= 0.8\textwidth]{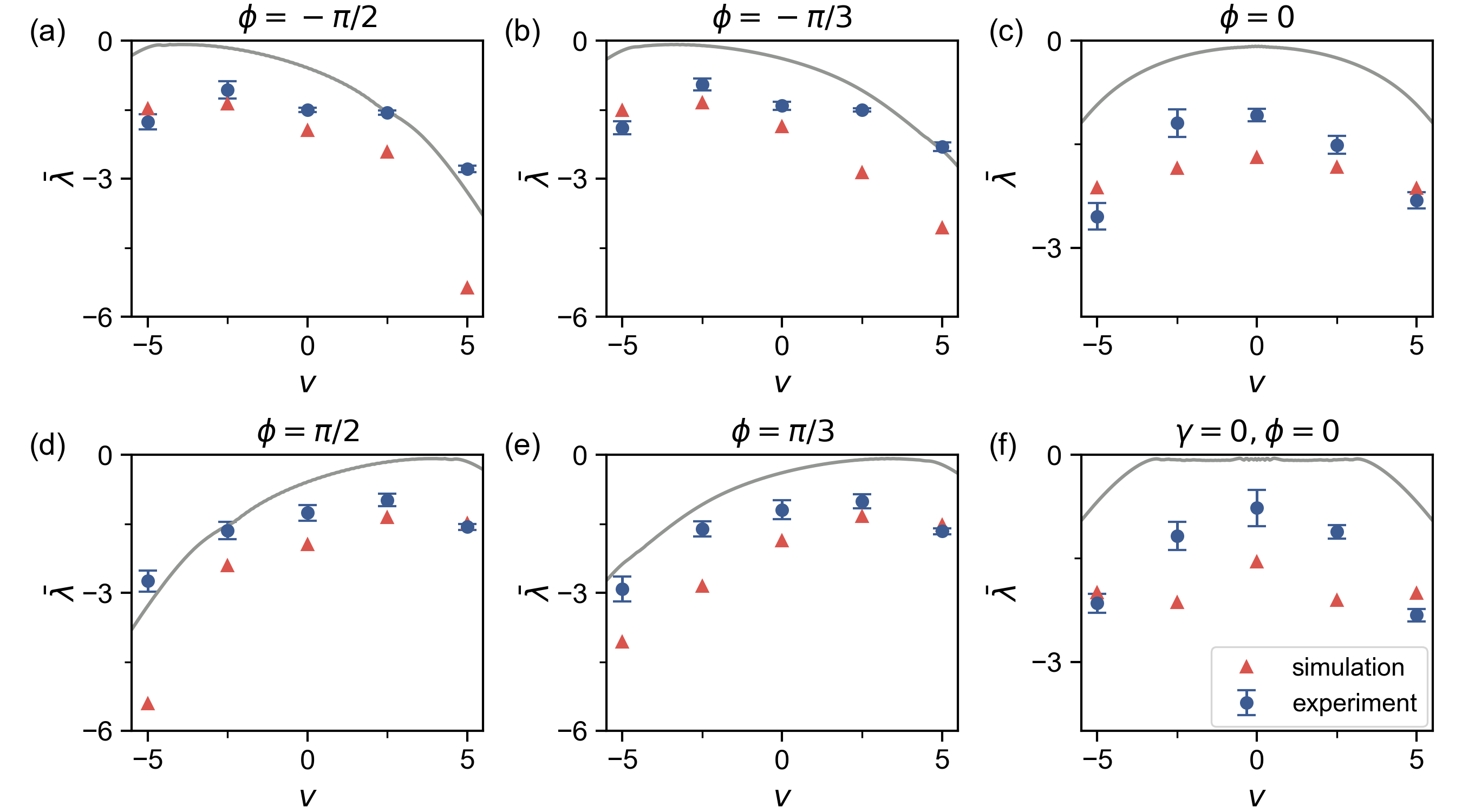}
\caption{\label{fig:Lp}
Signatures of the NHSE.
(a)(b)(c)(d)(e) Measured growth rate $\bar{\lambda}$ with a fixed  $\gamma=h\times 1.3(2)$\ kHz, and varying synthetic flux $\phi$. (f) $\bar{\lambda}(v)$ with $\phi = 0$ and $\gamma = 0$ and an evolution time of $0.4$\ ms. The blue dots are the experimental data, and the error bar represents SE of the mean. The solid lines are numerical results using the effective Hamiltonian for a lattice with $N=221$ unit cells. The red triangles are numerical simulations using the experimental parameters. For all panels, the hopping rates are  $\{J_s,~J_r,~J_p,~J_t\} = h \times \{1.13(2),~1.04(1),~0.62(4),~0.80(2)\}$\ kHz, and $\Delta = 0$.
}
\end{figure*}

{\it Dynamic signatures of NHSE.}
Hamiltonian~(1) hosts the NHSE, and can be topologically non-trivial under appropriate parameters~\cite{supp}.
In our setup, NHSE can be understood as arising from the non-reciprocity of the underlying dissipative AB rings~\cite{yanring}. As a result of the interplay of the synthetic flux and dissipation, a unidirectional
current emerges in the bulk, leading to the accumulation of population at boundaries under OBC, and forming the basis for the dynamic detection of NHSE. {Note that such a directional bulk current is also closely related to the spectral topology of the Hamiltonian~\cite{fangchenskin,supp}.}

We begin by experimentally confirming the presence of the directional bulk current. In Fig.~\ref{fig:typical}, we fix the synthetic flux $\phi=-\pi/2$, and show typical bulk dynamics, in terms of the time-evolution of the momentum-space
atomic distribution, for condensates initialized in the states $|n=0,a\rangle$ and $|n=0,b\rangle$, respectively. We focus on the condensate dynamics up to $t  =0.7$ ms, when atoms have not yet evolved to the edges. As demonstrated in Fig.~\ref{fig:typical}(a)(b), the condensate propagates to the left in the presence of dissipation, regardless of the initial state. By
contrast, in the Hermitian case [Fig.~\ref{fig:typical}(c)(d)], the direction of propagation is dependent on the initial sublattice state. The observation above indicates the presence of a non-reciprocal bulk flow with the onset of dissipation.

For a more quantitative characterization, we define a growth rate~\cite{stefano,lyaexp}
\begin{equation}
    \lambda(v,t)=  \frac{\log|\psi(n,t)|}{t},
\end{equation}
where $\psi(n,t)$ is the time-evolved wave function on the unit cell $n=vt$, and $v$ is the shift velocity.
Under NHSE, the growth rate peaks at a finite shift velocity $v_m$, after a finite time of evolution. Intuitively, $v_m$ reflects the speed of propagation for the peak of a wave packet initialized at $n=0$. Thus, a finite $v_m$ is a clear indication of the direction and strength of the bulk flow.

To ensure that the initial-state dependence of the condensate propagation in Fig.~\ref{fig:typical} is averaged out, we measure the mean growth rate $\bar{\lambda}=(\lambda_a+\lambda_b)/2$,
where $\lambda_{a,b}$ are the corresponding growth rates for condensates initialized in states $|n=0,a\rangle$ and $|n=0,b\rangle$, respectively.

In Fig.~\ref{fig:Lp}, we show measurements of $\bar{\lambda}$ for an evolution of $t=0.4$\ ms, with varying flux $\phi$. Under a finite flux, all measured growth rates peak at finite shift
velocities, where the sign of $v_m$ indicates the direction of the bulk flow. Under OBC, the flow direction is consistent with that of the boundary where eigen wavefunctions accumulate~\cite{supp}.
These measurements are in sharp contrast to the Hermitian case of Fig.~\ref{fig:Lp}(f), where $\bar{\lambda}$ peaks at $v_m=0$, and is symmetric with respect to the peak. Note that the small asymmetry with regard
to $v=0$, as observed in Fig.~\ref{fig:Lp}(c), comes from a remnant flux due to experimental imperfections.
Given its dependence on the flux and loss $\gamma$, the observed directional flow serves as a clear signature for the NHSE in our implemented model.

\begin{figure*}[tbp]
\centering
\includegraphics[width= 0.8\textwidth]{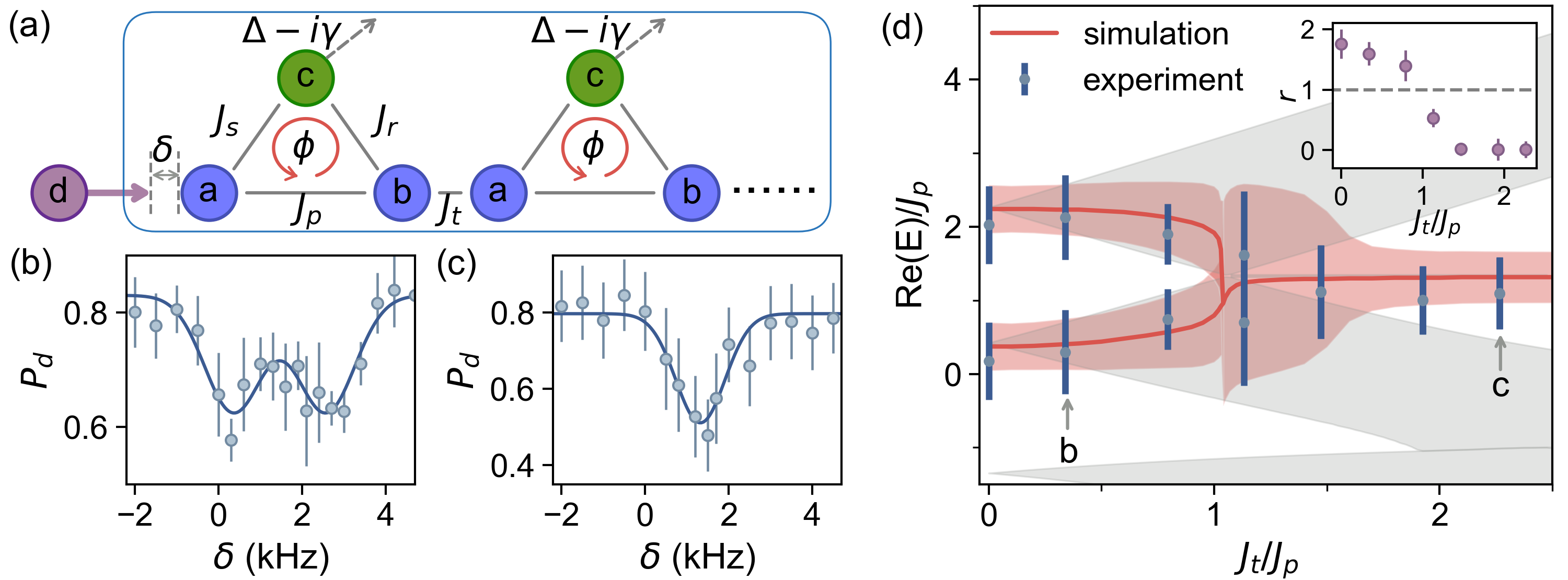}
\caption{\label{fig:energy}
Topological edge states through the Bragg spectroscopy.
(a) Schematic illustration of the Bragg spectroscopy. (b)(c) show typical experimental data (dots) of the Bragg spectrum, with $J_t/J_p=0.34$ in (b) and $J_t/J_p=2.27$ in (c). The solid lines are from the double-Gaussian fit.  Other experimental parameters are $\{J_s,~J_r,~J_p\}= h\times \{1.08(2),~1.02(3),~1.21(3)\}$\ kHz, $\gamma=h\times 1.0(1)$\ kHz, $\phi = -\pi/2$. An energy offset $\Delta=-h\times 2.0(3)$\ kHz on site $|n,c\rangle$ is also imposed to reduce its impact on the topological gap.
(d) Real components of the eigenspectrum (grey shaded), spectroscopic measurements (blue strips), and numerical simulations (red curves and shade). The band center is shifted from zero due to the light shift of the four-photon process~\cite{supp}.
The red curves and shade respectively indicate the valley location and width of the numerically simulated spectra.
The blue strips are determined by fitting the experimental data: the length of each strip corresponds to $2w$, and the blue dots indicate valley locations, calculated from $\delta_0$ and $\Delta\delta$ (see main text for definition).
The insert of (d) shows the distance-to-width ratio $r$ with increasing $J_t/J_p$, where the dashed horizontal line indicates the location of the phase transition with $r=1$.
Error bars show the SE of the mean.
}
\end{figure*}

{\it Resolving topological edge states.}
A unique challenge for non-Hermitian topological models with NHSE is the detection of topological phase transitions. In topological models without NHSE, the transition can be probed by detecting
topological edge states. However, this becomes difficult under NHSE, when all eigen wavefunctions are localized at the boundary along with the topological edge states. Here we adopt a Bragg spectroscopy with simultaneous spectral and spatial resolution to detect the topological edge states.

As shown in Fig.~\ref{fig:energy}(a), we prepare the condensate in an auxiliary momentum-lattice site, labeled $|d\rangle$ and encoded in the $|F=1,~m_F=0\rangle$ manifold. We then transfer atoms into the
dissipative AB chain by switching on a weak coupling ($\sim h\times 0.2$\ kHz) between site $|d\rangle$ and one edge of the lattice for a duration of $1$~ms. We measure the probability $P_d$ for atoms to remain in $|d\rangle$, while we vary the detuning $\delta$ of the coupling. Here $P_d$ is obtained by
dividing the atom number of $|d\rangle$ at the end of the Bragg coupling with the initial atom number. {Note that during the $1$ ms probe time, the edge-most three unit cells become occupied~\cite{supp}. Since the average localization length under the NHSE is of the same order [for instance $\sim 3.56$ under the parameters of Fig.~\ref{fig:energy}(b)]~\cite{supp}, the measurement is expected to reveal the global spectral information.}

In Fig.~\ref{fig:energy}(b)(c), we show typical spectroscopic data in the topologically trivial and non-trivial phases, respectively. A double-valley structure in the topologically trivial phase indicates the
presence of two bands, where state $|d\rangle$ is resonantly coupled to the bulk states. In comparison, only a single valley exists in the topologically non-trivial phase, where $|d\rangle$ is resonantly coupled
to the topological edge states. The valley location indicates the real eigenenergy component of the edge states. {While the real energy gap between the two topological edge states does not close in our finite-size system, the single valley appears when the gap becomes smaller than the experimental resolution.}

We fit the measured spectra using a double-Gaussian function
\begin{equation}\label{Eq:1}
P_d=A\left[e^{-\frac{(\delta-\delta_0)^2}{2w^2}}+e^{-\frac{(\delta-\delta_0-\Delta \delta)^2}{2w^2}}\right],
\end{equation}
In Fig.~\ref{fig:energy}(d), we show the fitted widths (half the length of the blue strips) and valley locations (blue dots) from experimental measurements. While the experimental results are consistent with those from numerical simulations (red curves and shade), they suggest the presence of a topological transition point, as a double-valley structure changes into a single-valley one. 
More specifically, we define a distance-to-width ratio $r=\Delta\delta/2w$, so that the transition occurs at $r=1$. As shown in the inset of Fig.~\ref{fig:energy}(d), $r$ undergoes a rapid change and drops from $\sim 1.5$ to $0$ around $J_t/J_p\sim 1$, indicative of a gap-closing transition in the thermodynamic limit. Theoretically, the topological phase transition is predicted by the non-Bloch band theory to occur at $J_t/J_p\sim 1.04$~\cite{supp}. By contrast, the Bloch winding number is not quantized close to the transition. Our observation is, therefore, consistent with the non-Bloch bulk-boundary correspondence~\cite{WZ1}.

To achieve a better resolution with narrower valleys, a weaker probe coupling (between $|d\rangle$ and the chain) and a longer probe time are desirable. Experimentally, however, these parameters are constrained by the decoherence time of the system, and the inhomogeneous broadening due to the interplay of trapping potential and interaction. As it is, the width of each valley is typically a few hundred Hz, which sets the resolution of the Bragg spectroscopy. Finally, in our experiment, the injected atoms propagate over three unit cells during the $1$\ ms probe time~\cite{supp}, which proves sufficient for the spectroscopy.

{\it Discussion.}
We report the first experimental identification of dynamic signatures of the NHSE in a Bose-Einstein condensate, by engineering a dissipative AB chain in the synthetic dimensions of hyperfine and momentum states of ultracold atoms. While we experimentally characterize the NHSE and its consequences using a finite-size system of five unit cells, it is desirable to further increase the experimentally achievable system size, which would reduce the finite-size effects~\cite{supp}, enable longer probe time and better resolutions. For this purpose, better phase locking of the coupling lasers and an initial condensate with more atom numbers are necessary.

For future studies, it would be fascinating to introduce the Feshbach resonance to the dissipative AB chain, where the inter-atomic interactions can be widely tuned \cite{Xie2020,Wang2022}. Our experiment thus paves the way toward a systematic study of NHSE in the many-body setting of cold atoms. Given the flexible control over atoms in momentum lattices, our scheme also has far-reaching implications for the immediate future. For instance, in light of recent experiments on localized states in momentum lattices~\cite{Meier2018,yanlocal}, an experimental characterization of the interplay between disorder, topology and NHSE is readily accessible based on our setup. Our work thus sheds new light on the study of quantum open systems.


\begin{acknowledgments}
{\it Acknowledgement:-}
B. Y. acknowledge the support from the National Key Research and Development Program of China under Grant No.2018YFA0307200, the National Natural Science Foundation of China under Grant No. 12074337, Natural
Science Foundation of Zhejiang province under Grant No. LR21A040002, Zhejiang Province Plan for Science and technology No. 2020C01019 and the Fundamental Research Funds for the Central Universities under
No. 2020XZZX002-05 and 2021FZZX001-02. W. Y. acknowledges support from the National Natural Science Foundation of China under Grant Nos. 11974331, the National Key Research and Development Program of China under
Grant Nos. 2017YFA0304100. B. G. acknowledges support by the U.S. Air Force Office of Scientific Research under Grant No. FA9550-21-1-0246.
\end{acknowledgments}

\bibliographystyle{apsrev4-2}
\bibliography{coldatomskinexp202205}

\begin{thebibliography}{49}%
\makeatletter
\providecommand \@ifxundefined [1]{%
 \@ifx{#1\undefined}
}%
\providecommand \@ifnum [1]{%
 \ifnum #1\expandafter \@firstoftwo
 \else \expandafter \@secondoftwo
 \fi
}%
\providecommand \@ifx [1]{%
 \ifx #1\expandafter \@firstoftwo
 \else \expandafter \@secondoftwo
 \fi
}%
\providecommand \natexlab [1]{#1}%
\providecommand \enquote  [1]{``#1''}%
\providecommand \bibnamefont  [1]{#1}%
\providecommand \bibfnamefont [1]{#1}%
\providecommand \citenamefont [1]{#1}%
\providecommand \href@noop [0]{\@secondoftwo}%
\providecommand \href [0]{\begingroup \@sanitize@url \@href}%
\providecommand \@href[1]{\@@startlink{#1}\@@href}%
\providecommand \@@href[1]{\endgroup#1\@@endlink}%
\providecommand \@sanitize@url [0]{\catcode `\\12\catcode `\$12\catcode
  `\&12\catcode `\#12\catcode `\^12\catcode `\_12\catcode `\%12\relax}%
\providecommand \@@startlink[1]{}%
\providecommand \@@endlink[0]{}%
\providecommand \url  [0]{\begingroup\@sanitize@url \@url }%
\providecommand \@url [1]{\endgroup\@href {#1}{\urlprefix }}%
\providecommand \urlprefix  [0]{URL }%
\providecommand \Eprint [0]{\href }%
\providecommand \doibase [0]{https://doi.org/}%
\providecommand \selectlanguage [0]{\@gobble}%
\providecommand \bibinfo  [0]{\@secondoftwo}%
\providecommand \bibfield  [0]{\@secondoftwo}%
\providecommand \translation [1]{[#1]}%
\providecommand \BibitemOpen [0]{}%
\providecommand \bibitemStop [0]{}%
\providecommand \bibitemNoStop [0]{.\EOS\space}%
\providecommand \EOS [0]{\spacefactor3000\relax}%
\providecommand \BibitemShut  [1]{\csname bibitem#1\endcsname}%
\let\auto@bib@innerbib\@empty
\bibitem [{\citenamefont {Carmichael}(1993)}]{QJ}%
  \BibitemOpen
  \bibfield  {author} {\bibinfo {author} {\bibfnamefont {H.~J.}\ \bibnamefont
  {Carmichael}},\ }\href {https://doi.org/10.1103/PhysRevLett.70.2273}
  {\bibfield  {journal} {\bibinfo  {journal} {Phys. Rev. Lett.}\ }\textbf
  {\bibinfo {volume} {70}},\ \bibinfo {pages} {2273} (\bibinfo {year}
  {1993})}\BibitemShut {NoStop}%
\bibitem [{\citenamefont {Ashida}\ \emph {et~al.}(2020)\citenamefont {Ashida},
  \citenamefont {Gong},\ and\ \citenamefont {Ueda}}]{uedareview}%
  \BibitemOpen
  \bibfield  {author} {\bibinfo {author} {\bibfnamefont {Y.}~\bibnamefont
  {Ashida}}, \bibinfo {author} {\bibfnamefont {Z.}~\bibnamefont {Gong}},\ and\
  \bibinfo {author} {\bibfnamefont {M.}~\bibnamefont {Ueda}},\ }\href
  {https://doi.org/10.1080/00018732.2021.1876991} {\bibfield  {journal}
  {\bibinfo  {journal} {Advances in Physics}\ }\textbf {\bibinfo {volume}
  {69}},\ \bibinfo {pages} {249} (\bibinfo {year} {2020})}\BibitemShut
  {NoStop}%
\bibitem [{\citenamefont {Bender}(2007)}]{ptbender}%
  \BibitemOpen
  \bibfield  {author} {\bibinfo {author} {\bibfnamefont {C.~M.}\ \bibnamefont
  {Bender}},\ }\href {https://doi.org/10.1088/0034-4885/70/6/r03} {\bibfield
  {journal} {\bibinfo  {journal} {Reports on Progress in Physics}\ }\textbf
  {\bibinfo {volume} {70}},\ \bibinfo {pages} {947} (\bibinfo {year}
  {2007})}\BibitemShut {NoStop}%
\bibitem [{\citenamefont {El-Ganainy}\ \emph {et~al.}(2018)\citenamefont
  {El-Ganainy}, \citenamefont {Makris}, \citenamefont {Khajavikhan},
  \citenamefont {Musslimani}, \citenamefont {Rotter},\ and\ \citenamefont
  {Christodoulides}}]{ptreview1}%
  \BibitemOpen
  \bibfield  {author} {\bibinfo {author} {\bibfnamefont {R.}~\bibnamefont
  {El-Ganainy}}, \bibinfo {author} {\bibfnamefont {K.~G.}\ \bibnamefont
  {Makris}}, \bibinfo {author} {\bibfnamefont {M.}~\bibnamefont {Khajavikhan}},
  \bibinfo {author} {\bibfnamefont {Z.~H.}\ \bibnamefont {Musslimani}},
  \bibinfo {author} {\bibfnamefont {S.}~\bibnamefont {Rotter}},\ and\ \bibinfo
  {author} {\bibfnamefont {D.~N.}\ \bibnamefont {Christodoulides}},\ }\href
  {https://doi.org/10.1038/nphys4323} {\bibfield  {journal} {\bibinfo
  {journal} {Nature Physics}\ }\textbf {\bibinfo {volume} {14}},\ \bibinfo
  {pages} {11} (\bibinfo {year} {2018})}\BibitemShut {NoStop}%
\bibitem [{\citenamefont {{\"O}zdemir}\ \emph {et~al.}(2019)\citenamefont
  {{\"O}zdemir}, \citenamefont {Rotter}, \citenamefont {Nori},\ and\
  \citenamefont {Yang}}]{ptreview2}%
  \BibitemOpen
  \bibfield  {author} {\bibinfo {author} {\bibfnamefont {{\c{S}}.~K.}\
  \bibnamefont {{\"O}zdemir}}, \bibinfo {author} {\bibfnamefont
  {S.}~\bibnamefont {Rotter}}, \bibinfo {author} {\bibfnamefont
  {F.}~\bibnamefont {Nori}},\ and\ \bibinfo {author} {\bibfnamefont
  {L.}~\bibnamefont {Yang}},\ }\href
  {https://doi.org/10.1038/s41563-019-0304-9} {\bibfield  {journal} {\bibinfo
  {journal} {Nature materials}\ }\textbf {\bibinfo {volume} {18}},\ \bibinfo
  {pages} {783} (\bibinfo {year} {2019})}\BibitemShut {NoStop}%
\bibitem [{\citenamefont {Miri}\ and\ \citenamefont {Alù}(2019)}]{ptreview3}%
  \BibitemOpen
  \bibfield  {author} {\bibinfo {author} {\bibfnamefont {M.-A.}\ \bibnamefont
  {Miri}}\ and\ \bibinfo {author} {\bibfnamefont {A.}~\bibnamefont {Alù}},\
  }\href {https://doi.org/10.1126/science.aar7709} {\bibfield  {journal}
  {\bibinfo  {journal} {Science}\ }\textbf {\bibinfo {volume} {363}},\ \bibinfo
  {pages} {eaar7709} (\bibinfo {year} {2019})}\BibitemShut {NoStop}%
\bibitem [{\citenamefont {Fruchart}\ \emph {et~al.}(2021)\citenamefont
  {Fruchart}, \citenamefont {Hanai}, \citenamefont {Littlewood},\ and\
  \citenamefont {Vitelli}}]{nonrecNature}%
  \BibitemOpen
  \bibfield  {author} {\bibinfo {author} {\bibfnamefont {M.}~\bibnamefont
  {Fruchart}}, \bibinfo {author} {\bibfnamefont {R.}~\bibnamefont {Hanai}},
  \bibinfo {author} {\bibfnamefont {P.~B.}\ \bibnamefont {Littlewood}},\ and\
  \bibinfo {author} {\bibfnamefont {V.}~\bibnamefont {Vitelli}},\ }\href
  {https://doi.org/10.1038/s41586-021-03375-9} {\bibfield  {journal} {\bibinfo
  {journal} {Nature}\ }\textbf {\bibinfo {volume} {592}},\ \bibinfo {pages}
  {363} (\bibinfo {year} {2021})}\BibitemShut {NoStop}%
\bibitem [{\citenamefont {Kawabata}\ \emph {et~al.}(2019)\citenamefont
  {Kawabata}, \citenamefont {Shiozaki}, \citenamefont {Ueda},\ and\
  \citenamefont {Sato}}]{nonHtopo1}%
  \BibitemOpen
  \bibfield  {author} {\bibinfo {author} {\bibfnamefont {K.}~\bibnamefont
  {Kawabata}}, \bibinfo {author} {\bibfnamefont {K.}~\bibnamefont {Shiozaki}},
  \bibinfo {author} {\bibfnamefont {M.}~\bibnamefont {Ueda}},\ and\ \bibinfo
  {author} {\bibfnamefont {M.}~\bibnamefont {Sato}},\ }\href
  {https://doi.org/10.1103/PhysRevX.9.041015} {\bibfield  {journal} {\bibinfo
  {journal} {Phys. Rev. X}\ }\textbf {\bibinfo {volume} {9}},\ \bibinfo {pages}
  {041015} (\bibinfo {year} {2019})}\BibitemShut {NoStop}%
\bibitem [{\citenamefont {Zhou}\ and\ \citenamefont {Lee}(2019)}]{nonHtopo2}%
  \BibitemOpen
  \bibfield  {author} {\bibinfo {author} {\bibfnamefont {H.}~\bibnamefont
  {Zhou}}\ and\ \bibinfo {author} {\bibfnamefont {J.~Y.}\ \bibnamefont {Lee}},\
  }\href {https://doi.org/10.1103/PhysRevB.99.235112} {\bibfield  {journal}
  {\bibinfo  {journal} {Phys. Rev. B}\ }\textbf {\bibinfo {volume} {99}},\
  \bibinfo {pages} {235112} (\bibinfo {year} {2019})}\BibitemShut {NoStop}%
\bibitem [{\citenamefont {Yao}\ and\ \citenamefont {Wang}(2018)}]{WZ1}%
  \BibitemOpen
  \bibfield  {author} {\bibinfo {author} {\bibfnamefont {S.}~\bibnamefont
  {Yao}}\ and\ \bibinfo {author} {\bibfnamefont {Z.}~\bibnamefont {Wang}},\
  }\href {https://doi.org/10.1103/PhysRevLett.121.086803} {\bibfield  {journal}
  {\bibinfo  {journal} {Phys. Rev. Lett.}\ }\textbf {\bibinfo {volume} {121}},\
  \bibinfo {pages} {086803} (\bibinfo {year} {2018})}\BibitemShut {NoStop}%
\bibitem [{\citenamefont {Yokomizo}\ and\ \citenamefont
  {Murakami}(2019)}]{murakami}%
  \BibitemOpen
  \bibfield  {author} {\bibinfo {author} {\bibfnamefont {K.}~\bibnamefont
  {Yokomizo}}\ and\ \bibinfo {author} {\bibfnamefont {S.}~\bibnamefont
  {Murakami}},\ }\href {https://doi.org/10.1103/PhysRevLett.123.066404}
  {\bibfield  {journal} {\bibinfo  {journal} {Phys. Rev. Lett.}\ }\textbf
  {\bibinfo {volume} {123}},\ \bibinfo {pages} {066404} (\bibinfo {year}
  {2019})}\BibitemShut {NoStop}%
\bibitem [{\citenamefont {Yao}\ \emph {et~al.}(2018)\citenamefont {Yao},
  \citenamefont {Song},\ and\ \citenamefont {Wang}}]{WZ2}%
  \BibitemOpen
  \bibfield  {author} {\bibinfo {author} {\bibfnamefont {S.}~\bibnamefont
  {Yao}}, \bibinfo {author} {\bibfnamefont {F.}~\bibnamefont {Song}},\ and\
  \bibinfo {author} {\bibfnamefont {Z.}~\bibnamefont {Wang}},\ }\href
  {https://doi.org/10.1103/PhysRevLett.121.136802} {\bibfield  {journal}
  {\bibinfo  {journal} {Phys. Rev. Lett.}\ }\textbf {\bibinfo {volume} {121}},\
  \bibinfo {pages} {136802} (\bibinfo {year} {2018})}\BibitemShut {NoStop}%
\bibitem [{\citenamefont {Lee}\ and\ \citenamefont
  {Thomale}(2019)}]{ThomalePRB}%
  \BibitemOpen
  \bibfield  {author} {\bibinfo {author} {\bibfnamefont {C.~H.}\ \bibnamefont
  {Lee}}\ and\ \bibinfo {author} {\bibfnamefont {R.}~\bibnamefont {Thomale}},\
  }\href {https://doi.org/10.1103/PhysRevB.99.201103} {\bibfield  {journal}
  {\bibinfo  {journal} {Phys. Rev. B}\ }\textbf {\bibinfo {volume} {99}},\
  \bibinfo {pages} {201103} (\bibinfo {year} {2019})}\BibitemShut {NoStop}%
\bibitem [{\citenamefont {Kunst}\ \emph {et~al.}(2018)\citenamefont {Kunst},
  \citenamefont {Edvardsson}, \citenamefont {Budich},\ and\ \citenamefont
  {Bergholtz}}]{Budich}%
  \BibitemOpen
  \bibfield  {author} {\bibinfo {author} {\bibfnamefont {F.~K.}\ \bibnamefont
  {Kunst}}, \bibinfo {author} {\bibfnamefont {E.}~\bibnamefont {Edvardsson}},
  \bibinfo {author} {\bibfnamefont {J.~C.}\ \bibnamefont {Budich}},\ and\
  \bibinfo {author} {\bibfnamefont {E.~J.}\ \bibnamefont {Bergholtz}},\ }\href
  {https://doi.org/10.1103/PhysRevLett.121.026808} {\bibfield  {journal}
  {\bibinfo  {journal} {Phys. Rev. Lett.}\ }\textbf {\bibinfo {volume} {121}},\
  \bibinfo {pages} {026808} (\bibinfo {year} {2018})}\BibitemShut {NoStop}%
\bibitem [{\citenamefont {McDonald}\ \emph {et~al.}(2018)\citenamefont
  {McDonald}, \citenamefont {Pereg-Barnea},\ and\ \citenamefont
  {Clerk}}]{mcdonald}%
  \BibitemOpen
  \bibfield  {author} {\bibinfo {author} {\bibfnamefont {A.}~\bibnamefont
  {McDonald}}, \bibinfo {author} {\bibfnamefont {T.}~\bibnamefont
  {Pereg-Barnea}},\ and\ \bibinfo {author} {\bibfnamefont {A.~A.}\ \bibnamefont
  {Clerk}},\ }\href {https://doi.org/10.1103/PhysRevX.8.041031} {\bibfield
  {journal} {\bibinfo  {journal} {Phys. Rev. X}\ }\textbf {\bibinfo {volume}
  {8}},\ \bibinfo {pages} {041031} (\bibinfo {year} {2018})}\BibitemShut
  {NoStop}%
\bibitem [{\citenamefont {Martinez~Alvarez}\ \emph {et~al.}(2018)\citenamefont
  {Martinez~Alvarez}, \citenamefont {Barrios~Vargas},\ and\ \citenamefont
  {Foa~Torres}}]{alvarez}%
  \BibitemOpen
  \bibfield  {author} {\bibinfo {author} {\bibfnamefont {V.~M.}\ \bibnamefont
  {Martinez~Alvarez}}, \bibinfo {author} {\bibfnamefont {J.~E.}\ \bibnamefont
  {Barrios~Vargas}},\ and\ \bibinfo {author} {\bibfnamefont {L.~E.~F.}\
  \bibnamefont {Foa~Torres}},\ }\href
  {https://doi.org/10.1103/PhysRevB.97.121401} {\bibfield  {journal} {\bibinfo
  {journal} {Phys. Rev. B}\ }\textbf {\bibinfo {volume} {97}},\ \bibinfo
  {pages} {121401} (\bibinfo {year} {2018})}\BibitemShut {NoStop}%
\bibitem [{\citenamefont {Zhang}\ \emph {et~al.}(2020)\citenamefont {Zhang},
  \citenamefont {Yang},\ and\ \citenamefont {Fang}}]{fangchenskin}%
  \BibitemOpen
  \bibfield  {author} {\bibinfo {author} {\bibfnamefont {K.}~\bibnamefont
  {Zhang}}, \bibinfo {author} {\bibfnamefont {Z.}~\bibnamefont {Yang}},\ and\
  \bibinfo {author} {\bibfnamefont {C.}~\bibnamefont {Fang}},\ }\href
  {https://doi.org/10.1103/PhysRevLett.125.126402} {\bibfield  {journal}
  {\bibinfo  {journal} {Phys. Rev. Lett.}\ }\textbf {\bibinfo {volume} {125}},\
  \bibinfo {pages} {126402} (\bibinfo {year} {2020})}\BibitemShut {NoStop}%
\bibitem [{\citenamefont {Okuma}\ \emph {et~al.}(2020)\citenamefont {Okuma},
  \citenamefont {Kawabata}, \citenamefont {Shiozaki},\ and\ \citenamefont
  {Sato}}]{kawabataskin}%
  \BibitemOpen
  \bibfield  {author} {\bibinfo {author} {\bibfnamefont {N.}~\bibnamefont
  {Okuma}}, \bibinfo {author} {\bibfnamefont {K.}~\bibnamefont {Kawabata}},
  \bibinfo {author} {\bibfnamefont {K.}~\bibnamefont {Shiozaki}},\ and\
  \bibinfo {author} {\bibfnamefont {M.}~\bibnamefont {Sato}},\ }\href
  {https://doi.org/10.1103/PhysRevLett.124.086801} {\bibfield  {journal}
  {\bibinfo  {journal} {Phys. Rev. Lett.}\ }\textbf {\bibinfo {volume} {124}},\
  \bibinfo {pages} {086801} (\bibinfo {year} {2020})}\BibitemShut {NoStop}%
\bibitem [{\citenamefont {Longhi}(2019)}]{stefano}%
  \BibitemOpen
  \bibfield  {author} {\bibinfo {author} {\bibfnamefont {S.}~\bibnamefont
  {Longhi}},\ }\href {https://doi.org/10.1103/PhysRevResearch.1.023013}
  {\bibfield  {journal} {\bibinfo  {journal} {Phys. Rev. Research}\ }\textbf
  {\bibinfo {volume} {1}},\ \bibinfo {pages} {023013} (\bibinfo {year}
  {2019})}\BibitemShut {NoStop}%
\bibitem [{\citenamefont {Yang}\ \emph {et~al.}(2020)\citenamefont {Yang},
  \citenamefont {Zhang}, \citenamefont {Fang},\ and\ \citenamefont
  {Hu}}]{yzsgbz}%
  \BibitemOpen
  \bibfield  {author} {\bibinfo {author} {\bibfnamefont {Z.}~\bibnamefont
  {Yang}}, \bibinfo {author} {\bibfnamefont {K.}~\bibnamefont {Zhang}},
  \bibinfo {author} {\bibfnamefont {C.}~\bibnamefont {Fang}},\ and\ \bibinfo
  {author} {\bibfnamefont {J.}~\bibnamefont {Hu}},\ }\href
  {https://doi.org/10.1103/PhysRevLett.125.226402} {\bibfield  {journal}
  {\bibinfo  {journal} {Phys. Rev. Lett.}\ }\textbf {\bibinfo {volume} {125}},\
  \bibinfo {pages} {226402} (\bibinfo {year} {2020})}\BibitemShut {NoStop}%
\bibitem [{\citenamefont {Deng}\ and\ \citenamefont {Yi}(2019)}]{tianshu}%
  \BibitemOpen
  \bibfield  {author} {\bibinfo {author} {\bibfnamefont {T.-S.}\ \bibnamefont
  {Deng}}\ and\ \bibinfo {author} {\bibfnamefont {W.}~\bibnamefont {Yi}},\
  }\href {https://doi.org/10.1103/PhysRevB.100.035102} {\bibfield  {journal}
  {\bibinfo  {journal} {Phys. Rev. B}\ }\textbf {\bibinfo {volume} {100}},\
  \bibinfo {pages} {035102} (\bibinfo {year} {2019})}\BibitemShut {NoStop}%
\bibitem [{\citenamefont {Li}\ \emph {et~al.}(2020)\citenamefont {Li},
  \citenamefont {Lee}, \citenamefont {Mu},\ and\ \citenamefont {Gong}}]{lli}%
  \BibitemOpen
  \bibfield  {author} {\bibinfo {author} {\bibfnamefont {L.}~\bibnamefont
  {Li}}, \bibinfo {author} {\bibfnamefont {C.~H.}\ \bibnamefont {Lee}},
  \bibinfo {author} {\bibfnamefont {S.}~\bibnamefont {Mu}},\ and\ \bibinfo
  {author} {\bibfnamefont {J.}~\bibnamefont {Gong}},\ }\href
  {https://doi.org/10.1038/s41467-020-18917-4} {\bibfield  {journal} {\bibinfo
  {journal} {Nature communications}\ }\textbf {\bibinfo {volume} {11}},\
  \bibinfo {pages} {1} (\bibinfo {year} {2020})}\BibitemShut {NoStop}%
\bibitem [{\citenamefont {Zhou}\ \emph {et~al.}(2021)\citenamefont {Zhou},
  \citenamefont {Li}, \citenamefont {Yi},\ and\ \citenamefont
  {Cui}}]{Zhou2021}%
  \BibitemOpen
  \bibfield  {author} {\bibinfo {author} {\bibfnamefont {L.}~\bibnamefont
  {Zhou}}, \bibinfo {author} {\bibfnamefont {H.}~\bibnamefont {Li}}, \bibinfo
  {author} {\bibfnamefont {W.}~\bibnamefont {Yi}},\ and\ \bibinfo {author}
  {\bibfnamefont {X.}~\bibnamefont {Cui}},\ }\href@noop {} {\bibfield
  {journal} {\bibinfo  {journal} {arXiv:2111.04196}\ } (\bibinfo {year}
  {2021})}\BibitemShut {NoStop}%
\bibitem [{\citenamefont {Guo}\ \emph {et~al.}(2021)\citenamefont {Guo},
  \citenamefont {Dong}, \citenamefont {Zhang}, \citenamefont {Hu},\ and\
  \citenamefont {Yang}}]{Guo2021}%
  \BibitemOpen
  \bibfield  {author} {\bibinfo {author} {\bibfnamefont {S.}~\bibnamefont
  {Guo}}, \bibinfo {author} {\bibfnamefont {C.}~\bibnamefont {Dong}}, \bibinfo
  {author} {\bibfnamefont {F.}~\bibnamefont {Zhang}}, \bibinfo {author}
  {\bibfnamefont {J.}~\bibnamefont {Hu}},\ and\ \bibinfo {author}
  {\bibfnamefont {Z.}~\bibnamefont {Yang}},\ }\href@noop {} {\bibfield
  {journal} {\bibinfo  {journal} {arXiv:2111.04220}\ } (\bibinfo {year}
  {2021})}\BibitemShut {NoStop}%
\bibitem [{\citenamefont {Li}\ \emph {et~al.}(2021)\citenamefont {Li},
  \citenamefont {Sun}, \citenamefont {Zhang},\ and\ \citenamefont
  {Yi}}]{tianyuquench}%
  \BibitemOpen
  \bibfield  {author} {\bibinfo {author} {\bibfnamefont {T.}~\bibnamefont
  {Li}}, \bibinfo {author} {\bibfnamefont {J.-Z.}\ \bibnamefont {Sun}},
  \bibinfo {author} {\bibfnamefont {Y.-S.}\ \bibnamefont {Zhang}},\ and\
  \bibinfo {author} {\bibfnamefont {W.}~\bibnamefont {Yi}},\ }\href
  {https://doi.org/10.1103/PhysRevResearch.3.023022} {\bibfield  {journal}
  {\bibinfo  {journal} {Phys. Rev. Research}\ }\textbf {\bibinfo {volume}
  {3}},\ \bibinfo {pages} {023022} (\bibinfo {year} {2021})}\BibitemShut
  {NoStop}%
\bibitem [{\citenamefont {Song}\ \emph {et~al.}(2019)\citenamefont {Song},
  \citenamefont {Yao},\ and\ \citenamefont {Wang}}]{wzopen}%
  \BibitemOpen
  \bibfield  {author} {\bibinfo {author} {\bibfnamefont {F.}~\bibnamefont
  {Song}}, \bibinfo {author} {\bibfnamefont {S.}~\bibnamefont {Yao}},\ and\
  \bibinfo {author} {\bibfnamefont {Z.}~\bibnamefont {Wang}},\ }\href
  {https://doi.org/10.1103/PhysRevLett.123.170401} {\bibfield  {journal}
  {\bibinfo  {journal} {Phys. Rev. Lett.}\ }\textbf {\bibinfo {volume} {123}},\
  \bibinfo {pages} {170401} (\bibinfo {year} {2019})}\BibitemShut {NoStop}%
\bibitem [{\citenamefont {Longhi}(2020)}]{stefanoopen}%
  \BibitemOpen
  \bibfield  {author} {\bibinfo {author} {\bibfnamefont {S.}~\bibnamefont
  {Longhi}},\ }\href {https://doi.org/10.1103/PhysRevB.102.201103} {\bibfield
  {journal} {\bibinfo  {journal} {Phys. Rev. B}\ }\textbf {\bibinfo {volume}
  {102}},\ \bibinfo {pages} {201103} (\bibinfo {year} {2020})}\BibitemShut
  {NoStop}%
\bibitem [{\citenamefont {Helbig}\ \emph {et~al.}(2020)\citenamefont {Helbig},
  \citenamefont {Hofmann}, \citenamefont {Imhof}, \citenamefont {Abdelghany},
  \citenamefont {Kiessling}, \citenamefont {Molenkamp}, \citenamefont {Lee},
  \citenamefont {Szameit}, \citenamefont {Greiter},\ and\ \citenamefont
  {Thomale}}]{teskin}%
  \BibitemOpen
  \bibfield  {author} {\bibinfo {author} {\bibfnamefont {T.}~\bibnamefont
  {Helbig}}, \bibinfo {author} {\bibfnamefont {T.}~\bibnamefont {Hofmann}},
  \bibinfo {author} {\bibfnamefont {S.}~\bibnamefont {Imhof}}, \bibinfo
  {author} {\bibfnamefont {M.}~\bibnamefont {Abdelghany}}, \bibinfo {author}
  {\bibfnamefont {T.}~\bibnamefont {Kiessling}}, \bibinfo {author}
  {\bibfnamefont {L.}~\bibnamefont {Molenkamp}}, \bibinfo {author}
  {\bibfnamefont {C.}~\bibnamefont {Lee}}, \bibinfo {author} {\bibfnamefont
  {A.}~\bibnamefont {Szameit}}, \bibinfo {author} {\bibfnamefont
  {M.}~\bibnamefont {Greiter}},\ and\ \bibinfo {author} {\bibfnamefont
  {R.}~\bibnamefont {Thomale}},\ }\href
  {https://doi.org/10.1038/s41567-020-0922-9} {\bibfield  {journal} {\bibinfo
  {journal} {Nature Physics}\ }\textbf {\bibinfo {volume} {16}},\ \bibinfo
  {pages} {747} (\bibinfo {year} {2020})}\BibitemShut {NoStop}%
\bibitem [{\citenamefont {Xiao}\ \emph
  {et~al.}(2020{\natexlab{a}})\citenamefont {Xiao}, \citenamefont {Deng},
  \citenamefont {Wang}, \citenamefont {Zhu}, \citenamefont {Wang},
  \citenamefont {Yi},\ and\ \citenamefont {Xue}}]{photonskin}%
  \BibitemOpen
  \bibfield  {author} {\bibinfo {author} {\bibfnamefont {L.}~\bibnamefont
  {Xiao}}, \bibinfo {author} {\bibfnamefont {T.}~\bibnamefont {Deng}}, \bibinfo
  {author} {\bibfnamefont {K.}~\bibnamefont {Wang}}, \bibinfo {author}
  {\bibfnamefont {G.}~\bibnamefont {Zhu}}, \bibinfo {author} {\bibfnamefont
  {Z.}~\bibnamefont {Wang}}, \bibinfo {author} {\bibfnamefont {W.}~\bibnamefont
  {Yi}},\ and\ \bibinfo {author} {\bibfnamefont {P.}~\bibnamefont {Xue}},\
  }\href {https://doi.org/10.1038/s41567-020-0836-6} {\bibfield  {journal}
  {\bibinfo  {journal} {Nature Physics}\ }\textbf {\bibinfo {volume} {16}},\
  \bibinfo {pages} {761} (\bibinfo {year} {2020}{\natexlab{a}})}\BibitemShut
  {NoStop}%
\bibitem [{\citenamefont {Xiao}\ \emph
  {et~al.}(2021{\natexlab{a}})\citenamefont {Xiao}, \citenamefont {Deng},
  \citenamefont {Wang}, \citenamefont {Wang}, \citenamefont {Yi},\ and\
  \citenamefont {Xue}}]{XDW+21}%
  \BibitemOpen
  \bibfield  {author} {\bibinfo {author} {\bibfnamefont {L.}~\bibnamefont
  {Xiao}}, \bibinfo {author} {\bibfnamefont {T.}~\bibnamefont {Deng}}, \bibinfo
  {author} {\bibfnamefont {K.}~\bibnamefont {Wang}}, \bibinfo {author}
  {\bibfnamefont {Z.}~\bibnamefont {Wang}}, \bibinfo {author} {\bibfnamefont
  {W.}~\bibnamefont {Yi}},\ and\ \bibinfo {author} {\bibfnamefont
  {P.}~\bibnamefont {Xue}},\ }\href
  {https://doi.org/10.1103/PhysRevLett.126.230402} {\bibfield  {journal}
  {\bibinfo  {journal} {Phys. Rev. Lett.}\ }\textbf {\bibinfo {volume} {126}},\
  \bibinfo {pages} {230402} (\bibinfo {year} {2021}{\natexlab{a}})}\BibitemShut
  {NoStop}%
\bibitem [{\citenamefont {Ghatak}\ \emph {et~al.}(2020)\citenamefont {Ghatak},
  \citenamefont {Brandenbourger}, \citenamefont {van Wezel},\ and\
  \citenamefont {Coulais}}]{metaskin}%
  \BibitemOpen
  \bibfield  {author} {\bibinfo {author} {\bibfnamefont {A.}~\bibnamefont
  {Ghatak}}, \bibinfo {author} {\bibfnamefont {M.}~\bibnamefont
  {Brandenbourger}}, \bibinfo {author} {\bibfnamefont {J.}~\bibnamefont {van
  Wezel}},\ and\ \bibinfo {author} {\bibfnamefont {C.}~\bibnamefont
  {Coulais}},\ }\href {https://doi.org/10.1073/pnas.2010580117} {\bibfield
  {journal} {\bibinfo  {journal} {Proceedings of the National Academy of
  Sciences}\ }\textbf {\bibinfo {volume} {117}},\ \bibinfo {pages} {29561}
  (\bibinfo {year} {2020})}\BibitemShut {NoStop}%
\bibitem [{\citenamefont {Hofmann}\ \emph {et~al.}(2020)\citenamefont
  {Hofmann}, \citenamefont {Helbig}, \citenamefont {Schindler}, \citenamefont
  {Salgo}, \citenamefont {Brzezi\ifmmode~\acute{n}\else \'{n}\fi{}ska},
  \citenamefont {Greiter}, \citenamefont {Kiessling}, \citenamefont {Wolf},
  \citenamefont {Vollhardt}, \citenamefont {Kaba\ifmmode~\check{s}\else
  \v{s}\fi{}i}, \citenamefont {Lee}, \citenamefont {Bilu\ifmmode \check{s}\else
  \v{s}\fi{}i\ifmmode~\acute{c}\else \'{c}\fi{}}, \citenamefont {Thomale},\
  and\ \citenamefont {Neupert}}]{teskin2d}%
  \BibitemOpen
  \bibfield  {author} {\bibinfo {author} {\bibfnamefont {T.}~\bibnamefont
  {Hofmann}}, \bibinfo {author} {\bibfnamefont {T.}~\bibnamefont {Helbig}},
  \bibinfo {author} {\bibfnamefont {F.}~\bibnamefont {Schindler}}, \bibinfo
  {author} {\bibfnamefont {N.}~\bibnamefont {Salgo}}, \bibinfo {author}
  {\bibfnamefont {M.}~\bibnamefont {Brzezi\ifmmode~\acute{n}\else
  \'{n}\fi{}ska}}, \bibinfo {author} {\bibfnamefont {M.}~\bibnamefont
  {Greiter}}, \bibinfo {author} {\bibfnamefont {T.}~\bibnamefont {Kiessling}},
  \bibinfo {author} {\bibfnamefont {D.}~\bibnamefont {Wolf}}, \bibinfo {author}
  {\bibfnamefont {A.}~\bibnamefont {Vollhardt}}, \bibinfo {author}
  {\bibfnamefont {A.}~\bibnamefont {Kaba\ifmmode~\check{s}\else \v{s}\fi{}i}},
  \bibinfo {author} {\bibfnamefont {C.~H.}\ \bibnamefont {Lee}}, \bibinfo
  {author} {\bibfnamefont {A.}~\bibnamefont {Bilu\ifmmode \check{s}\else
  \v{s}\fi{}i\ifmmode~\acute{c}\else \'{c}\fi{}}}, \bibinfo {author}
  {\bibfnamefont {R.}~\bibnamefont {Thomale}},\ and\ \bibinfo {author}
  {\bibfnamefont {T.}~\bibnamefont {Neupert}},\ }\href
  {https://doi.org/10.1103/PhysRevResearch.2.023265} {\bibfield  {journal}
  {\bibinfo  {journal} {Phys. Rev. Research}\ }\textbf {\bibinfo {volume}
  {2}},\ \bibinfo {pages} {023265} (\bibinfo {year} {2020})}\BibitemShut
  {NoStop}%
\bibitem [{\citenamefont {Weidemann}\ \emph {et~al.}(2020)\citenamefont
  {Weidemann}, \citenamefont {Kremer}, \citenamefont {Helbig}, \citenamefont
  {Hofmann}, \citenamefont {Stegmaier}, \citenamefont {Greiter}, \citenamefont
  {Thomale},\ and\ \citenamefont {Szameit}}]{scienceskin}%
  \BibitemOpen
  \bibfield  {author} {\bibinfo {author} {\bibfnamefont {S.}~\bibnamefont
  {Weidemann}}, \bibinfo {author} {\bibfnamefont {M.}~\bibnamefont {Kremer}},
  \bibinfo {author} {\bibfnamefont {T.}~\bibnamefont {Helbig}}, \bibinfo
  {author} {\bibfnamefont {T.}~\bibnamefont {Hofmann}}, \bibinfo {author}
  {\bibfnamefont {A.}~\bibnamefont {Stegmaier}}, \bibinfo {author}
  {\bibfnamefont {M.}~\bibnamefont {Greiter}}, \bibinfo {author} {\bibfnamefont
  {R.}~\bibnamefont {Thomale}},\ and\ \bibinfo {author} {\bibfnamefont
  {A.}~\bibnamefont {Szameit}},\ }\href
  {https://doi.org/10.1126/science.aaz8727} {\bibfield  {journal} {\bibinfo
  {journal} {Science}\ }\textbf {\bibinfo {volume} {368}},\ \bibinfo {pages}
  {311} (\bibinfo {year} {2020})}\BibitemShut {NoStop}%
\bibitem [{\citenamefont {Meier}\ \emph {et~al.}(2016)\citenamefont {Meier},
  \citenamefont {An},\ and\ \citenamefont {Gadway}}]{Meier2016}%
  \BibitemOpen
  \bibfield  {author} {\bibinfo {author} {\bibfnamefont {E.~J.}\ \bibnamefont
  {Meier}}, \bibinfo {author} {\bibfnamefont {F.~A.}\ \bibnamefont {An}},\ and\
  \bibinfo {author} {\bibfnamefont {B.}~\bibnamefont {Gadway}},\ }\href
  {https://doi.org/10.1103/PhysRevA.93.051602} {\bibfield  {journal} {\bibinfo
  {journal} {Phys. Rev. A}\ }\textbf {\bibinfo {volume} {93}},\ \bibinfo
  {pages} {051602} (\bibinfo {year} {2016})}\BibitemShut {NoStop}%
\bibitem [{\citenamefont {Meier}\ \emph {et~al.}(2018)\citenamefont {Meier},
  \citenamefont {An}, \citenamefont {Dauphin}, \citenamefont {Maffei},
  \citenamefont {Massignan}, \citenamefont {Hughes},\ and\ \citenamefont
  {Gadway}}]{Meier2018}%
  \BibitemOpen
  \bibfield  {author} {\bibinfo {author} {\bibfnamefont {E.~J.}\ \bibnamefont
  {Meier}}, \bibinfo {author} {\bibfnamefont {F.~A.}\ \bibnamefont {An}},
  \bibinfo {author} {\bibfnamefont {A.}~\bibnamefont {Dauphin}}, \bibinfo
  {author} {\bibfnamefont {M.}~\bibnamefont {Maffei}}, \bibinfo {author}
  {\bibfnamefont {P.}~\bibnamefont {Massignan}}, \bibinfo {author}
  {\bibfnamefont {T.~L.}\ \bibnamefont {Hughes}},\ and\ \bibinfo {author}
  {\bibfnamefont {B.}~\bibnamefont {Gadway}},\ }\href
  {https://doi.org/10.1126/science.aat3406} {\bibfield  {journal} {\bibinfo
  {journal} {Science}\ }\textbf {\bibinfo {volume} {362}},\ \bibinfo {pages}
  {929} (\bibinfo {year} {2018})}\BibitemShut {NoStop}%
\bibitem [{\citenamefont {Lapp}\ \emph {et~al.}(2019)\citenamefont {Lapp},
  \citenamefont {Ang'ong'a}, \citenamefont {An},\ and\ \citenamefont
  {Gadway}}]{Lapp2019}%
  \BibitemOpen
  \bibfield  {author} {\bibinfo {author} {\bibfnamefont {S.}~\bibnamefont
  {Lapp}}, \bibinfo {author} {\bibfnamefont {J.}~\bibnamefont {Ang'ong'a}},
  \bibinfo {author} {\bibfnamefont {F.~A.}\ \bibnamefont {An}},\ and\ \bibinfo
  {author} {\bibfnamefont {B.}~\bibnamefont {Gadway}},\ }\href
  {https://doi.org/10.1088/1367-2630/ab1147} {\bibfield  {journal} {\bibinfo
  {journal} {New Journal of Physics}\ }\textbf {\bibinfo {volume} {21}},\
  \bibinfo {pages} {045006} (\bibinfo {year} {2019})}\BibitemShut {NoStop}%
\bibitem [{\citenamefont {Xie}\ \emph {et~al.}(2019)\citenamefont {Xie},
  \citenamefont {Gou}, \citenamefont {Xiao}, \citenamefont {Gadway},\ and\
  \citenamefont {Yan}}]{Xie2019}%
  \BibitemOpen
  \bibfield  {author} {\bibinfo {author} {\bibfnamefont {D.}~\bibnamefont
  {Xie}}, \bibinfo {author} {\bibfnamefont {W.}~\bibnamefont {Gou}}, \bibinfo
  {author} {\bibfnamefont {T.}~\bibnamefont {Xiao}}, \bibinfo {author}
  {\bibfnamefont {B.}~\bibnamefont {Gadway}},\ and\ \bibinfo {author}
  {\bibfnamefont {B.}~\bibnamefont {Yan}},\ }\href
  {https://doi.org/10.1038/s41534-019-0159-6} {\bibfield  {journal} {\bibinfo
  {journal} {npj Quantum Information}\ }\textbf {\bibinfo {volume} {5}},\
  \bibinfo {pages} {1} (\bibinfo {year} {2019})}\BibitemShut {NoStop}%
\bibitem [{\citenamefont {Xie}\ \emph {et~al.}(2018)\citenamefont {Xie},
  \citenamefont {Wang}, \citenamefont {Gou}, \citenamefont {Bu},\ and\
  \citenamefont {Yan}}]{Xie2018}%
  \BibitemOpen
  \bibfield  {author} {\bibinfo {author} {\bibfnamefont {D.}~\bibnamefont
  {Xie}}, \bibinfo {author} {\bibfnamefont {D.}~\bibnamefont {Wang}}, \bibinfo
  {author} {\bibfnamefont {W.}~\bibnamefont {Gou}}, \bibinfo {author}
  {\bibfnamefont {W.}~\bibnamefont {Bu}},\ and\ \bibinfo {author}
  {\bibfnamefont {B.}~\bibnamefont {Yan}},\ }\href
  {https://doi.org/10.1364/JOSAB.35.000500} {\bibfield  {journal} {\bibinfo
  {journal} {J. Opt. Soc. Am. B}\ }\textbf {\bibinfo {volume} {35}},\ \bibinfo
  {pages} {500} (\bibinfo {year} {2018})}\BibitemShut {NoStop}%
\bibitem [{\citenamefont {Gou}\ \emph {et~al.}(2020)\citenamefont {Gou},
  \citenamefont {Chen}, \citenamefont {Xie}, \citenamefont {Xiao},
  \citenamefont {Deng}, \citenamefont {Gadway}, \citenamefont {Yi},\ and\
  \citenamefont {Yan}}]{yanring}%
  \BibitemOpen
  \bibfield  {author} {\bibinfo {author} {\bibfnamefont {W.}~\bibnamefont
  {Gou}}, \bibinfo {author} {\bibfnamefont {T.}~\bibnamefont {Chen}}, \bibinfo
  {author} {\bibfnamefont {D.}~\bibnamefont {Xie}}, \bibinfo {author}
  {\bibfnamefont {T.}~\bibnamefont {Xiao}}, \bibinfo {author} {\bibfnamefont
  {T.-S.}\ \bibnamefont {Deng}}, \bibinfo {author} {\bibfnamefont
  {B.}~\bibnamefont {Gadway}}, \bibinfo {author} {\bibfnamefont
  {W.}~\bibnamefont {Yi}},\ and\ \bibinfo {author} {\bibfnamefont
  {B.}~\bibnamefont {Yan}},\ }\href
  {https://doi.org/10.1103/PhysRevLett.124.070402} {\bibfield  {journal}
  {\bibinfo  {journal} {Phys. Rev. Lett.}\ }\textbf {\bibinfo {volume} {124}},\
  \bibinfo {pages} {070402} (\bibinfo {year} {2020})}\BibitemShut {NoStop}%
\bibitem [{\citenamefont {Chen}\ \emph {et~al.}(2021)\citenamefont {Chen},
  \citenamefont {Gou}, \citenamefont {Xie}, \citenamefont {Xiao}, \citenamefont
  {Yi}, \citenamefont {Jing},\ and\ \citenamefont {Yan}}]{Chen2021}%
  \BibitemOpen
  \bibfield  {author} {\bibinfo {author} {\bibfnamefont {T.}~\bibnamefont
  {Chen}}, \bibinfo {author} {\bibfnamefont {W.}~\bibnamefont {Gou}}, \bibinfo
  {author} {\bibfnamefont {D.}~\bibnamefont {Xie}}, \bibinfo {author}
  {\bibfnamefont {T.}~\bibnamefont {Xiao}}, \bibinfo {author} {\bibfnamefont
  {W.}~\bibnamefont {Yi}}, \bibinfo {author} {\bibfnamefont {J.}~\bibnamefont
  {Jing}},\ and\ \bibinfo {author} {\bibfnamefont {B.}~\bibnamefont {Yan}},\
  }\href {https://doi.org/10.1038/s41534-021-00417-y} {\bibfield  {journal}
  {\bibinfo  {journal} {npj Quantum Information}\ }\textbf {\bibinfo {volume}
  {7}},\ \bibinfo {pages} {1} (\bibinfo {year} {2021})}\BibitemShut {NoStop}%
\bibitem [{sup()}]{supp}%
  \BibitemOpen
  \href@noop {} {}\bibinfo {note} {See Supplemental Materials for details,
  which includes Refs.
  \cite{PRx2014,epjd2020,WZ1,murakami,fangchenskin,kawabataskin,yanring,gberry1,gberry2}.}\BibitemShut
  {Stop}%
\bibitem [{\citenamefont {Lin}\ \emph {et~al.}(2021)\citenamefont {Lin},
  \citenamefont {Li}, \citenamefont {Xiao}, \citenamefont {Wang}, \citenamefont
  {Yi},\ and\ \citenamefont {Xue}}]{lyaexp}%
  \BibitemOpen
  \bibfield  {author} {\bibinfo {author} {\bibfnamefont {Q.}~\bibnamefont
  {Lin}}, \bibinfo {author} {\bibfnamefont {T.}~\bibnamefont {Li}}, \bibinfo
  {author} {\bibfnamefont {L.}~\bibnamefont {Xiao}}, \bibinfo {author}
  {\bibfnamefont {K.}~\bibnamefont {Wang}}, \bibinfo {author} {\bibfnamefont
  {W.}~\bibnamefont {Yi}},\ and\ \bibinfo {author} {\bibfnamefont
  {P.}~\bibnamefont {Xue}},\ }\href {https://arxiv.org/abs/2108.01109}
  {\bibfield  {journal} {\bibinfo  {journal} {arXiv:2108.01097}\ } (\bibinfo
  {year} {2021})}\BibitemShut {NoStop}%
\bibitem [{\citenamefont {Xie}\ \emph {et~al.}(2020)\citenamefont {Xie},
  \citenamefont {Deng}, \citenamefont {Xiao}, \citenamefont {Gou},
  \citenamefont {Chen}, \citenamefont {Yi},\ and\ \citenamefont
  {Yan}}]{Xie2020}%
  \BibitemOpen
  \bibfield  {author} {\bibinfo {author} {\bibfnamefont {D.}~\bibnamefont
  {Xie}}, \bibinfo {author} {\bibfnamefont {T.-S.}\ \bibnamefont {Deng}},
  \bibinfo {author} {\bibfnamefont {T.}~\bibnamefont {Xiao}}, \bibinfo {author}
  {\bibfnamefont {W.}~\bibnamefont {Gou}}, \bibinfo {author} {\bibfnamefont
  {T.}~\bibnamefont {Chen}}, \bibinfo {author} {\bibfnamefont {W.}~\bibnamefont
  {Yi}},\ and\ \bibinfo {author} {\bibfnamefont {B.}~\bibnamefont {Yan}},\
  }\href {https://doi.org/10.1103/PhysRevLett.124.050502} {\bibfield  {journal}
  {\bibinfo  {journal} {Phys. Rev. Lett.}\ }\textbf {\bibinfo {volume} {124}},\
  \bibinfo {pages} {050502} (\bibinfo {year} {2020})}\BibitemShut {NoStop}%
\bibitem [{\citenamefont {Wang}\ \emph {et~al.}(2022)\citenamefont {Wang},
  \citenamefont {Zhang}, \citenamefont {Li}, \citenamefont {Wu}, \citenamefont
  {Liu}, \citenamefont {Mei}, \citenamefont {Hu}, \citenamefont {Xiao},
  \citenamefont {Ma}, \citenamefont {Chin},\ and\ \citenamefont
  {Jia}}]{Wang2022}%
  \BibitemOpen
  \bibfield  {author} {\bibinfo {author} {\bibfnamefont {Y.}~\bibnamefont
  {Wang}}, \bibinfo {author} {\bibfnamefont {J.-H.}\ \bibnamefont {Zhang}},
  \bibinfo {author} {\bibfnamefont {Y.}~\bibnamefont {Li}}, \bibinfo {author}
  {\bibfnamefont {J.}~\bibnamefont {Wu}}, \bibinfo {author} {\bibfnamefont
  {W.}~\bibnamefont {Liu}}, \bibinfo {author} {\bibfnamefont {F.}~\bibnamefont
  {Mei}}, \bibinfo {author} {\bibfnamefont {Y.}~\bibnamefont {Hu}}, \bibinfo
  {author} {\bibfnamefont {L.}~\bibnamefont {Xiao}}, \bibinfo {author}
  {\bibfnamefont {J.}~\bibnamefont {Ma}}, \bibinfo {author} {\bibfnamefont
  {C.}~\bibnamefont {Chin}},\ and\ \bibinfo {author} {\bibfnamefont
  {S.}~\bibnamefont {Jia}},\ }\href@noop {} {\bibfield  {journal} {\bibinfo
  {journal} {arXiv:2204.12730}\ } (\bibinfo {year} {2022})}\BibitemShut
  {NoStop}%
\bibitem [{\citenamefont {Xiao}\ \emph
  {et~al.}(2021{\natexlab{b}})\citenamefont {Xiao}, \citenamefont {Xie},
  \citenamefont {Dong}, \citenamefont {Chen}, \citenamefont {Yi},\ and\
  \citenamefont {Yan}}]{yanlocal}%
  \BibitemOpen
  \bibfield  {author} {\bibinfo {author} {\bibfnamefont {T.}~\bibnamefont
  {Xiao}}, \bibinfo {author} {\bibfnamefont {D.}~\bibnamefont {Xie}}, \bibinfo
  {author} {\bibfnamefont {Z.}~\bibnamefont {Dong}}, \bibinfo {author}
  {\bibfnamefont {T.}~\bibnamefont {Chen}}, \bibinfo {author} {\bibfnamefont
  {W.}~\bibnamefont {Yi}},\ and\ \bibinfo {author} {\bibfnamefont
  {B.}~\bibnamefont {Yan}},\ }\href
  {https://doi.org/https://doi.org/10.1016/j.scib.2021.07.025} {\bibfield
  {journal} {\bibinfo  {journal} {Science Bulletin}\ }\textbf {\bibinfo
  {volume} {66}},\ \bibinfo {pages} {2175} (\bibinfo {year}
  {2021}{\natexlab{b}})}\BibitemShut {NoStop}%
\bibitem [{\citenamefont {Goldman}\ and\ \citenamefont
  {Dalibard}(2014)}]{PRx2014}%
  \BibitemOpen
  \bibfield  {author} {\bibinfo {author} {\bibfnamefont {N.}~\bibnamefont
  {Goldman}}\ and\ \bibinfo {author} {\bibfnamefont {J.}~\bibnamefont
  {Dalibard}},\ }\href {https://doi.org/10.1103/PhysRevX.4.031027} {\bibfield
  {journal} {\bibinfo  {journal} {Phys. Rev. X}\ }\textbf {\bibinfo {volume}
  {4}},\ \bibinfo {pages} {031027} (\bibinfo {year} {2014})}\BibitemShut
  {NoStop}%
\bibitem [{\citenamefont {Xiao}\ \emph
  {et~al.}(2020{\natexlab{b}})\citenamefont {Xiao}, \citenamefont {Xie},
  \citenamefont {Gou}, \citenamefont {Deng}, \citenamefont {Yi},\ and\
  \citenamefont {Yan}}]{epjd2020}%
  \BibitemOpen
  \bibfield  {author} {\bibinfo {author} {\bibfnamefont {T.}~\bibnamefont
  {Xiao}}, \bibinfo {author} {\bibfnamefont {D.}~\bibnamefont {Xie}}, \bibinfo
  {author} {\bibfnamefont {W.}~\bibnamefont {Gou}}, \bibinfo {author}
  {\bibfnamefont {T.-S.}\ \bibnamefont {Deng}}, \bibinfo {author}
  {\bibfnamefont {W.}~\bibnamefont {Yi}},\ and\ \bibinfo {author}
  {\bibfnamefont {B.}~\bibnamefont {Yan}},\ }\href
  {https://doi.org/10.1140/epjd/e2020-10019-6} {\bibfield  {journal} {\bibinfo
  {journal} {Eur. Phys. J. D.}\ }\textbf {\bibinfo {volume} {74}},\ \bibinfo
  {pages} {152} (\bibinfo {year} {2020}{\natexlab{b}})}\BibitemShut {NoStop}%
\bibitem [{\citenamefont {Liang}\ and\ \citenamefont {Huang}(2013)}]{gberry1}%
  \BibitemOpen
  \bibfield  {author} {\bibinfo {author} {\bibfnamefont {S.-D.}\ \bibnamefont
  {Liang}}\ and\ \bibinfo {author} {\bibfnamefont {G.-Y.}\ \bibnamefont
  {Huang}},\ }\href {https://doi.org/10.1103/PhysRevA.87.012118} {\bibfield
  {journal} {\bibinfo  {journal} {Phys. Rev. A}\ }\textbf {\bibinfo {volume}
  {87}},\ \bibinfo {pages} {012118} (\bibinfo {year} {2013})}\BibitemShut
  {NoStop}%
\bibitem [{\citenamefont {Lieu}(2018)}]{gberry2}%
  \BibitemOpen
  \bibfield  {author} {\bibinfo {author} {\bibfnamefont {S.}~\bibnamefont
  {Lieu}},\ }\href {https://doi.org/10.1103/PhysRevB.97.045106} {\bibfield
  {journal} {\bibinfo  {journal} {Phys. Rev. B}\ }\textbf {\bibinfo {volume}
  {97}},\ \bibinfo {pages} {045106} (\bibinfo {year} {2018})}\BibitemShut
  {NoStop}%
\end{thebibliography}%

\end{document}